\documentclass[aps,twocolumn,showpacs,preprintnumbers]{revtex4}

\usepackage{graphicx}  
\usepackage{dcolumn}   
\usepackage{bm}        
\usepackage{subfigure}

\newcommand{\gamdot}{\dot{\gamma}}

\begin{document}

\title{Oscillatory Motion of Rising Bubbles in Wormlike Micellar Fluids \\ with Different Microstructures}

\author{Nestor Z. Handzy and Andrew Belmonte}

\affiliation {\it The W.~G.~Pritchard Laboratories, Department of Mathematics,
Penn State University, University Park, PA 16802}

\date{October 15, 2003; submitted January 10, 2003}

\begin{abstract}  

Previous observations of the nontransient oscillations of rising bubbles and
falling spheres in wormlike micellar fluids were limited to a single surfactant
system. We present an extensive survey of rising bubbles in another system, an aqueous solution of cetylpyridinium chloride and sodium
salicylate, with and without NaCl, across a range of  concentrations and
temperatures.  Two different types of oscillation are seen in different
concentration ranges, each with its own temperature dependence. Rheological data
allows for the identification of these different hydrodynamic states with
different fluid microstructures.

\end{abstract}


\maketitle

Fluids are often broadly categorized as either Newtonian or non-Newtonian,
according to whether the Navier-Stokes equation does or does not accurately
describe the fluid's motion.  Newtonian fluids consist of molecules small
enough to be approximated by point masses (such as air and water), while
non-Newtonian consist of larger structures (such as polymers).  This increased
size affords greater degrees of freedom which leads to macroscopic
viscoelasticity \cite{bird}.  Fluids consisting of self-assembling cylindrical
(or wormlike) micelles resemble polymeric fluids on the microscopic scale, with
the added feature that their length distribution is determined by aggregation
kinetics; micelles continually break and reform \cite{israelachvilibook,
gelbart, cates90a}. The molecular level physics is clearly more complicated for
such fluids, and one can reasonably expect this will introduce a new set of
flow properties to the class of non-Newtonian fluids.

Several novel results have been reported in both viscometric flows
\cite{rehage88, shikata88, grand, lerouge00, liuandpine} and hydrodynamic
(non-viscometric) flows \cite{smolka01, ab00, anand01} of various wormlike
micellar fluids.  Among these, a falling pendant drop at the bottom of a thin
filament has been observed to slow to a complete stop before the filament
suddenly ruptures \cite{smolka01}; this rupture has been confirmed in other
configurations \cite{john}.  Also, rising air bubbles and falling solid spheres
have been observed to oscillate without reaching a terminal velocity
\cite{ab00,anand01}.  While in polymer solutions a rising bubble displays a
sharp non-axisymmetric cusp, which remains unchanged during a steady rise
\cite{bird, liu95}, in wormlike micellar fluids the cusp periodically extends to
a sharp point, then retracts to a blunt edge (see Fig.1).  It is likely that
these new dynamics are a manifestation of the reversible scission reactions of
the micelles, though the precise mechanism may be quite complicated.  Another
type of fluid is known to include scission-like reactions at its microscale -
associating polymer solutions - and falling sphere oscillations are reported in
that system as well \cite{mollinger, weidman}.

\begin{figure}
\begin{center}
\includegraphics[width=7.0cm]{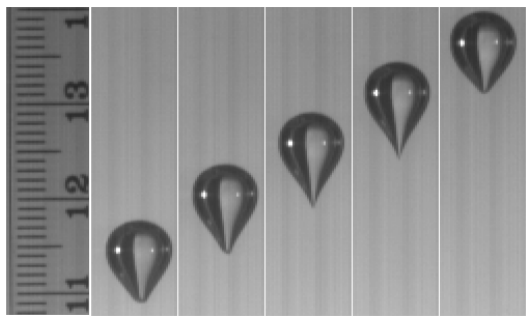}
\caption{Cusp shape change of an oscillating bubble rising through a 
15mM 
CPCl/NaSal (weight fraction $\varphi = 0.5 \%$) solution at $T= 
37.5^{\circ}$C. The scale at left is marked in centimeters.  Interval 
between pictures: 0.05 s.}
\label{Type1}
\end{center}
\end{figure}

\begin{figure}
\begin{center}

\includegraphics[width=7.8cm]{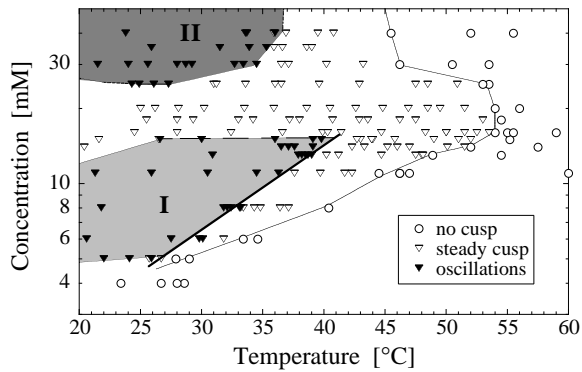}

\caption{\label{CTplane1} Temperature and concentration phase diagram  for the dynamics of rising
bubbles in equimolar CPCl/NaSal, showing two distinct regions of oscillating  behavior (shaded)
labelled as I and II. The thick sloped line marking
the temperature boundary for type I oscillations is an isoline of Eqn. 1.}

\end{center}
\end{figure}

In this Letter we report a detailed survey of the oscillatory motion of rising
bubbles in a wormlike micellar fluid.  To study the role of the aggregation
kinetics of the micelles, both concentration and fluid
temperature were varied over a wide range, unlike previous studies \cite{ab00,
anand01}.  As these parameters change, different micellar architectures are
possible - from short linear or branched micelles to crosslinked networks
\cite{appell92, schmitt94,  in99}.  Four
different dynamics were seen: Newtonian behavior at high temperatures, standard
polymeric behavior, and two distinct oscillating responses occurring in
different concentration ranges (Fig.\ref{CTplane1}).  We performed steady
rheology experiments to identify the fluid microstate, and found that
transitions in the equilibrium structure match transitions in bubble dynamics.
Critical temperature bounds were also found, which can be interpreted as a
minimum length of micelle required for oscillations to occur, so that for
certain concentrations the fluid may be tuned with temperature to make bubbles
either oscillate, rise with a stable cusp, or rise as bubbles in a Newtonian
fluid.

Non-transient oscillating bubbles were first observed in aqueous solutions of
cetyltrimethylammonium bromide (CTAB) and sodium salicylate (NaSal) \cite{ab00};
here we use another familiar system, cetylpyridinium chloride (CPCl) and NaSal,
with the fixed ratio [NaSal]/[CPCl] = 1 (except for the NaCl experiments
described below).  Our choice for this ratio gives a solution of flexible
cylindrical micelles analogous to polymer chains \cite{rehage88}.     All
experimental results are for a single rising  bubble, with volumes ranging from
14 mm$^3$ to 110 mm$^3$.  The bubble is injected into the fluid at the bottom
of a temperature controlled  plastic cylinder (31 cm height, 5 cm diameter)
using a syringe with a  long stainless steel tube (inner diameter of either 1.0
mm or 1.5  mm).   The ratio of the horizontal diameter of the bubble $d$ to the
cylinder diameter $D$ is $d/D \leq 0.14$, and we have also checked that bubbles
oscillate for $d/D \simeq 0.02$. Typical sizes of the Reynolds number (inertia)
are $Re \simeq 10^{-2} - 5$, while the Deborah number (elasticity) ranged from
$De \simeq 1 - 500$, similar to values seen for oscillating bubbles and spheres
in CTAB/NaSal \cite{ab00, anand01}.  Rheological data were taken with a
Rheometrics RFS-III controlled rate of strain rheometer with circulating fluid
temperature bath, using a stainless steel Couette geometry.  Video images were
made with a Kodak high speed CCD camera.


We have studied concentrations from 4 to 40 mM (weight fractions $0.13\% \leq
\varphi \leq 1.3 \%$).  From 5 to 15 mM, bubbles have a cusp which oscillates
in length and changes shape (Fig.1).  While rising, the cusp lengthens (frames
1-4 in Fig.1) during which the velocity as measured at the top of the bubble is
nearly constant.  At the apex of the extension, the tail abruptly retracts and
the bubble jumps upward (frames 4-5).  After this it slows to a nearly
constant velocity until the cycle repeats. Typically, bubbles begin to
oscillate within 10 seconds of their formation, with a similar time between
jumps. These clearly visible oscillations, which we call type I, are apparently
not a transient effect; we have observed their persistence for rise distances
over one meter, during which there were more than 30 oscillations in $\sim$ 35
s (for 8 mM CPCl/NaSal), and velocities more than doubled during a jump. This
is the typical type I behavior in the temperature ranges shown in
Fig.\ref{CTplane1}, consistent with the oscillations observed
in CTAB/NaSal \cite{ab00}.

Above a critical temperature in the type I range (Fig.\ref{CTplane1}), bubbles
have a sharp cusp which does not change in length (``steady cusp'' in
Fig.\ref{CTplane1}) and velocities smoothly reach a steady state; the
oscillatory instability vanishes.  By $60^{\circ}$C, the solutions appear
Newtonian (``no cusp'' in Fig.\ref{CTplane1}): large enough bubbles are
ellipsoidal and undergo the well-known side-to-side oscillations \cite{hart57}.
In this temperature range, the micelles must be predominantly spherical or
short rigid rods \cite{ben-shaul}.


\begin{figure}
\begin{center}
\includegraphics[width=3.0in]{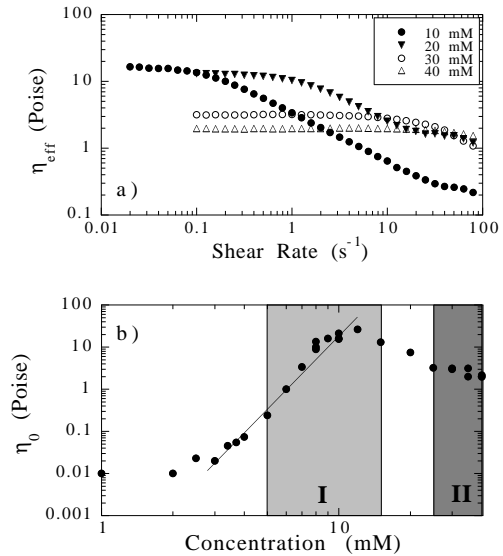}

\caption{\label{zero} Rheology of equimolar CPCl/NaSal at $30^{\circ}$C: a)
effective viscosity versus shear rate for 10, 20, 30, and 40 mM; b) zero shear
viscosity, $\eta_0$ as a function of concentration. The shaded regions mark the
concentration ranges of bubble oscillations, and the sloped line starting at 3
mM corresponds to $\eta_0\sim \varphi^{5.8}$.}

\end{center}
\end{figure}

Upon increasing the concentration from 15 mM to 16 mM ($\varphi= 0.5\%$ to
$0.53\%$), \emph{all oscillations cease}.  Bubbles rise steadily with a stable
cusp for concentrations up to 20 mM; there are no temperatures at which oscillations occur (Fig.\ref{CTplane1}).  It
is striking  that a rising bubble would be so affected by so slight a
concentration change. Note the transition temperature from polymeric to
Newtonian behavior reaches a maximum at this concentration (Fig.\ref{CTplane1}).

More surprising is the re-emergence of oscillations for concentrations from 25
mM to 40 mM ($0.8 \% \leq \varphi \leq 1.3\%$).  Yet here the oscillations,
which we call type II, are visibly different from type I oscillations; the shape change involves the entire bubble, whereas in type I
it seems restricted to the tail.  A type II oscillation begins with a
constriction in width near the top, while the whole bubble lengthens.  This
constriction then travels downward, as if the bubble were squeezing through a
hoop.  Defining $w_{max}$ to be the maximum (relaxed) width (at its waist),
$w_{min}$ to be its most constricted width, and $\Delta w= w_{max} - w_{min}$,
we found $\Delta w /w_{max} = 0.13$ in type II fluids and $0.04$ in type I. 
Length extensions, however, were $ \Delta l / l_{min} = 0.25$ in type II and
$0.27$ in type I; for bubbles rising in CTAB/NaSal, $\Delta l/l_{min} = 0.26$ 
\cite{ab00}. This previous study revealed only one type of oscillation, which from the shape dynamics we identify as type I \cite{ab00}.

Although wormlike micelles are comprised of surfactants, it is unlikely that
surface tension plays a role in the bubble oscillations, since falling rigid
spheres also oscillate \cite{anand01}.  Furthermore, our observation of two
oscillation types suggests that the viscoelastic character of the bulk fluid
is changing with concentration. We address this with rheology measurements,
controlling the shear strain rate $\dot{\gamma}$ and recording  the effective
viscosity, $\eta_{\text{eff}} = \sigma_{xy} / \dot{\gamma}$, where $\sigma_{xy}$
is  the steady shear stress.  Transient tests were also performed to ensure
that steady state  was achieved.  Shown in Fig.\ref{zero}a is
$\eta_{\text{eff}}$ vs.~$\dot{\gamma}$ at $30^{\circ}$C, for
fluids in the oscillating and non-oscillating concentration ranges. Most fluids
are shear thinning - $\eta_{\text{eff}}$  decreases at high enough
$\dot{\gamma}$ - however fluids below 5 mM shear thicken ($\eta_{\text{eff}}$
increases above $\eta_0$), typically requiring $\sim 100$ seconds to reach
steady state \cite{liuandpine}.  Bubbles tested in fluids below 5 mM showed no
oscillations or steady cusps. While there is a noticeable difference in the
rheology upon increasing from 20 to 30 mM,  30 and 40 mM (type II region) are
strikingly similar; they also have a constant viscosity over a broader $\gamdot$
range than the other fluids.  More interestingly, the zero shear  viscosity
$\eta_0$ (defined as the plateau value at low $\dot{\gamma}$) decreases with
concentration.

A more complete study of the dependence of $\eta_0$ on $\varphi$ indicates
transitions in fluid microstructure (Fig.\ref{zero}b). The transition at low concentration to rapid $\eta_0$ growth marks the overlap concentration $\varphi^*
\simeq 3$ mM, below which is the dilute regime. Above $\varphi^*$ is the semi-dilute state, in which micelles exist as individual entangled worms \cite{berret93}.  Here $\eta_0 \sim
\varphi^{5.8}$, close to the value $5.4$ associated with stress relaxation by reptation \cite{appell92, candau93}.  The semi-dilute regime ends
near $\varphi \simeq 12 - 15$ mM, followed by an extraordinary decrease
of $\eta_0$ with concentration, continuing to $\varphi
\simeq 30$ mM, after which $\eta_0$ varies weakly.


\begin{figure}
\begin{center}
\includegraphics[width=5.75cm]{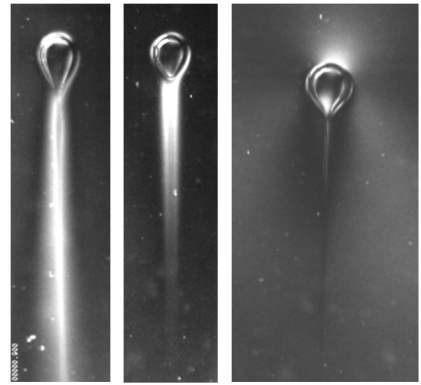}

\caption{\label{Biref} Birefringent images of the wake behind a bubble rising in CPCl/NaSal at $T=
24^{\circ}$C: a) 10 mM ; b) 20 mM ; c) 35 mM. Each image is 6.5 cm high. Reynolds and Deborah 
numbers for each image are: a) $Re \simeq 4.72$, $De \simeq 250$; b) $Re \simeq 0.02$, $De \simeq
6$; c) $Re \simeq 1.1$, $De \simeq 1.8$.}

\end{center}
\end{figure}


The dramatic change in the concentration dependence of $\eta_0$  indicates a
transition in the equilibrium fluid structure,  coinciding with the loss of
type I oscillations. It may be that the entangled  micelles in the semi-dilute
range have begun to fuse at entanglement points, forming a crosslinked network
\cite{appell92, candau93, drye, pincus!}.  The junction nodes are free to slide
along the micelles in such a formation, which would account for the decrease in
viscosity \cite{appell92}.   If the ratio of crosslinks to entanglement points
grows as $\eta_0$ decreases, then the new state would be fully formed where
$\eta_0$ stabilizes near $\sim 30$ mM ($1\%$), the start of type II oscillations.  Note that a crosslinked network state has also been proposed for
CTAB/NaSal at $\varphi \simeq 1\%$ \cite{lequeux97}.

Transitions in micellar morphology would naturally lead to transitions in 
mechanisms for stress relaxation, which should be observable in the stress field
around the rising bubble. These transitions are made evident with
birefringent visualization, shown in Fig.\ref{Biref} for the three concentration
regimes.  Optical birefringence is a well-known technique for visualizing stress
in non-Newtonian fluids  \cite{opticalannrev}, especially effective for wormlike
micellar fluids \cite{hu93}.  Fig.\ref{Biref}a (10 mM) shows the localization of
stress in the wake of a type I oscillating bubble.  This birefringent tail
mirrors the dynamics of the bubble's cusp (Fig.\ref{Type1}), and when it retracts, the birefringent tail suddenly disperses to the sides.  The
flow in the bubble wake is nearly in uniaxial extension \cite{harlen90}, and the
birefringent tail suggests a strong relation to a thin filament whose rupture
\cite{smolka01} may be related to the mechanism for oscillation. At 20 mM a shorter tail is seen (Fig.\ref{Biref}b), whose length remains
constant (like the bubble tail).  Fig.\ref{Biref}c corresponds to type II
oscillations (35 mM), in which three equally spaced birefringent bands indicate
a broader distribution of stress.  This pattern is essentially unaltered during an oscillation. Note that  $De$ decreases from type I to type II,
with an intermediate value for non-oscillating bubbles (Fig.4 caption), contradicting the implicit critical $De$ condition for oscillations in
\cite{anand01}.

The length distribution of the individual worms in the type I region
(semi-dilute) should depend on temperature and concentration. In equilibrium, 
the average length of a wormlike micelle $L_{0}$ can be described with the mean
field approximation \cite{cates90a, israelachvilibook, ben-shaul}: 
\begin{equation}  
L_{0}(\phi,T) \sim \sqrt {\phi} \;e^{E/2kT}   
\end{equation} 
where $\phi$ is the total amphiphile concentration, $k$ is Boltzmann's 
constant, $T$ is temperature, and $E$ is the scission energy required to break
one wormlike micelle into two.  Estimates based on light scattering measurements
for $E$ are on the order of $10kT$ \cite{cates90a, kwon, in99} - much lower than
the covalent bonds of polymers, yet large enough for some micelles to reach
appreciable lengths against thermal fluctuations \cite{cates90a}.  The flow near
the bubble may increase the equilibrium length, and there is evidence that
shear-induced structures (SIS) much larger than individual micelles form in such
flow in both CTAB/NaSal \cite{liuandpine, bolt97, skeller98} and CPCl/NaSal
\cite{wheeler, fischer}. We assume simply that for the type I region, $L_{0}$ is
the relevant quantity determining if the bubble will oscillate.  Specifically,
we attribute the oscillations to a breaking instability of the elongated
micelles or SIS in the wake, occurring only if $L_{0}$ exceeds some critical
value. Using Eqn.1, the solid line in Fig.\ref{CTplane1} corresponds to a
scission energy of $E= 1.01 \times 10^{-19}$ Joules, about $24kT$.

Type II oscillations have a much simpler dependence on temperature; they occur
only for  $T \lesssim T_c = 36^{\circ}$C (Fig.\ref{CTplane1}).  If crosslinks
have indeed formed for $T < T_c$, then the mean field length  is no longer an
appropriate quantity. The transition temperature $T_c$ suggests a critical
energy condition ($E_c \simeq 4.3 \times  10^{-21}$ Joules), which may
correspond to the energy difference between junctions and endcaps. Thus for $T >
T_c$, endcaps would dominate and the fluid is comprised of individual micelles
which can entangle like polymers.  This possible explanation is consistent with
both the rheology and the rising bubble dynamics.

The two different oscillations we have observed appear linked to two different
microstructures.  We extended our study to include the well
characterized ternary system CPCl-NaSal diluted in concentrated NaCl, with a
ratio [NaSal]/[CPCl] = 0.5 \cite{berret93, berret94, lerouge00}.  We tested
several concentrations \cite{conc} of these fluids for jumping bubbles in our
apparatus, at $T= 30^{\circ}$C, a temperature central to both type I and II
oscillations (Fig.\ref{CTplane1}). We observed no oscillations or any behavior
different from steadily rising bubbles in conventional polymeric fluids
\cite{liu95}. This classic ternary system is known to consist of entangled wormlike
micelles, which when taken with the observed scaling $\eta_0\sim
\varphi^{3.3}$, indicates that micellar breaking occurs on a shorter timescale
than reptation \cite{cates90a, berret93}; in contrast for our fluids
$\eta_0\sim \varphi^{5.8}$.  Evidently, the oscillatory motion of rising
bubbles is not a characteristic of fluids in this ``fast-breaking'' limit
\cite{cates90a,berret97}, and we conclude that type I oscillations are in the
``slow-breaking'' limit.

Our study of rising bubbles in various wormlike micellar fluids 
indicates that while scission reactions may be necessary for oscillations, there are other conditions. The discovery of a second type of
oscillation provides another example of how microscale dynamics and architecture
(entangled vs.~crosslinked) interact to produce macroscopic instabilities. It now seems that wormlike micellar fluids are the most generic complex fluids as far as rising bubbles are concerned, since all known material-dependent dynamics can occur at different temperatures or concentrations. More surprisingly, the rise of an air bubble has been shown to be extraordinarily sensitive to fluid microstructure.

We thank A.~Jayaraman, J.~T.~Jacobsen, P.~Olmsted, T.~Podgorski, and
L.~M.~Walker for valuable discussions, R.~Geist, D.~M.~Henderson, and
M.~C.~Sostarecz for experimental assistance, and the referees for constructive
comments.  AB acknowledges support from the A.~P.~Sloan Foundation and the
National Science Foundation (CAREER Award DMR-0094167).



\begin{thebibliography}{99}
\bibitem{conc} Concentrations tested were:  [CPCl] = 30, 40, 50, 60, 70, and 80 mM, in the system
[NaSal]/[CPCl] = 0.5 with 500 mM NaCl.
\end{thebibliography}
\end{document}